\documentclass[aps,preprint,superscriptaddress]{revtex4}

\usepackage{amsmath}

\begin{document}
\bibliographystyle{apsrev}


\title{Critical behaviour of the two-dimensional Ising susceptibility}



\author{W. P. Orrick}
\email[]{orrick@labri.u-bordeaux.fr}
\affiliation{Department of Mathematics \& Statistics,
The University of Melbourne,
Parkville, Vic. 3010, Australia}

\author{B. G. Nickel}
\email[]{BGN@physics.uoguelph.ca}
\affiliation{Department of Physics,
University of Guelph,
Guelph, Ontario, Canada N1G 2W1}

\author{A. J. Guttmann}
\email[]{tonyg@ms.unimelb.edu.au}
\affiliation{Department of Mathematics \& Statistics,
The University of Melbourne,
Parkville, Vic. 3010, Australia}

\author{J. H. H. Perk}
\email[]{perk@okstate.edu}
\affiliation{Department of Physics, Oklahoma State University,
Stillwater, Oklahoma 74078-3072, U.S.A.}

\date{\today}

\begin{abstract}
We report computations of the short-distance and the
long-distance (scaling) contributions to the
square-lattice Ising susceptibility in zero field close to $T_c$.
Both computations rely on the use of nonlinear partial difference
equations for the correlation functions. By summing the correlation
functions, we give an algorithm of complexity O$(N^6)$ for the determination
of the first $N$ series coefficients. Consequently, we have generated 
and analysed series of length several hundred terms, generated
in about 100 hours on an obsolete workstation.
In terms of a temperature variable, $\tau$, linear in
$T/T_c-1$, the short-distance terms are shown to have the form
$\tau^p(\ln|\tau|)^q$ with $p\ge q^2$.  
To $\rm{O}(\tau^{14})$ the long-distance part divided by the leading
$\tau^{-7/4}$  singularity
contains only integer powers of $\tau$.  The presence of irrelevant
variables in the scaling function is clearly evident, with
contributions of distinct character at leading
orders $|\tau|^{9/4}$ and $|\tau|^{17/4}$ being identified.
\end{abstract}

\maketitle

\section{Introduction}\label{sec:intro}

The two-dimensional Ising model has been
extremely useful as a testing ground for new theoretical ideas and methods in
the study of phase transitions and critical phenomena. 
Our present understanding is the
result of a series of dramatic developments spanning more than half a century,
starting with Onsager's exact computation of the free energy~\cite{Onsager44},
followed by Yang's derivation of the spontaneous magnetization~\cite{Yang52}
and by the work of many researchers on the correlation functions, including
Toeplitz determinantal formulae~\cite{MontrollPottsWard63}, exact expressions
for their behaviour at large separation~\cite{WMTB76}, and nonlinear partial
difference equations for their efficient
computation~\cite{McCoyWu80,Perk80,JimboMiwa80}, to mention only those
results which are used in the present work.  All results above apply
to the zero-field case. 
While an exact expression for the susceptibility as the sum of
two-point correlation functions over all separations~\cite{WMTB76}
exists, a useful closed form expression does not.
Moreover, as we discuss, there are strong
indications that the susceptibility has a natural boundary in the complex
plane~\cite{GuttmannEnting96,Nickel99}, a feature which rules out any
expression in terms of the ``standard'' functions of mathematical physics.

Nevertheless, it is desirable to obtain as detailed information about
the susceptibility as possible, not only because of its physical
importance, but also because of the significant role it plays in
ideas about scaling and the renormalization group. In the vicinity of
the ferromagnetic critical point at temperature $T=T_c$, the
susceptibility exhibits a singularity of the form
\begin{equation}
\beta^{-1}\chi_{\pm}=
C_{0_\pm}(2K_c\sqrt{2})^{7/4}|\tau|^{-7/4}F_{\pm}(\tau)+B(\tau).
\label{introequation}
\end{equation}
Here $\beta=(k_BT)^{-1}$, $\tau=\tfrac{1}{2}(s^{-1}-s)$, $s=\sinh 2K$ and
$2K_c=\ln(1+\sqrt{2})$ with $K=\beta J$ the
conventional Ising model coupling constant.  The scaling-amplitude functions
$F_{\pm}(\tau)$ are normalized to unity at $\tau=0.$  As a
consequence of the exact knowledge of the long-range correlations,
the coefficients $C_{0_\pm}$ were calculated
exactly~\cite{BarouchMcCoyWu73} in terms of the solution of a
Painlev{\'e} III equation.  Additionally, the leading behaviour of
both $F_{\pm}(\tau)$ was computed to be $1+\tfrac{1}{2}\tau$. The
antiferromagnetic susceptibility, on the other hand, is dominated by
the short-distance correlation functions and has leading singularity
$(\text{const}\times\tau\ln|\tau|)$. Such short-distance
``background'' terms are present as well in the ferromagnetic
susceptibility and are denoted by $B(\tau)$ in (\ref{introequation}). 
The leading amplitudes of the analytic and singular parts of
$B(\tau)$ were computed for a general wavevector dependent
susceptibility
in~\cite{KongAuYangPerk86a,Kong87}.

An analysis~\cite{GartenhausMcCullough88} of a 51 term
high-temperature series by means of differential approximants yielded two
further correction terms in the scaling-amplitude function $F_+$, with
numerical amplitudes close to rational values,
$\tfrac{5}{8}\tau^2+\tfrac{3}{16}\tau^3$, and confirmed that the same 
scaling-amplitude function is numerically consistent with the first 
11 terms in the
low-temperature expansion. These results agreed with the 
prediction~\cite{AharonyFisher80} that the corrections to
scaling are entirely due to the nonlinearity of the scaling fields and not to
the presence of irrelevant operators~\cite{AharonyFisher80}.
However, a recent analysis of 115 term high- and low-temperature
series~\cite{Nickel00} showed that this prediction appears to break down in the
amplitude of $\tau^4$. 

The study reported in this letter substantially improves on all the above
results. We extend the methods
of~\cite{KongAuYangPerk86a,Kong87} to compute both
antiferromagnetic and ferromagnetic background amplitudes on the
isotropic lattice
to ${\rm O}(\tau^{14}(\ln|\tau|)^3)$. All such terms are seen to be of the form
$\tau^p(\ln|\tau|)^q$ with $p\ge q^2$.
We simultaneously compute high-temperature series to order 323 and
low-temperature series to order 646 in 123 hours on a 500MHz DEC Alpha with
21164 processor running Maple$^{\text{TM}}$ V version 5.1.

We analyze these series by two independent methods, making use of the
computed background amplitudes and the known complex singularity
structure~\cite{Nickel99,Nickel00} to obtain the scaling-amplitude functions
$F_{\pm}$ to ${\rm O}(\tau^{14})$.

Several important conclusions can be drawn from our results. Firstly, only
pure integer powers of $\tau$ enter the scaling-amplitude functions
and no logarithmic terms are present.
Secondly, the high- and low-temperature scaling-amplitude 
functions are not equal to each other. The amplitudes start to differ
at ${\rm O}(\tau^6)$. Thirdly, the amplitudes of $\tau^4$ and $\tau^5$, which
are clearly rational, are not those predicted by simple two-variable
scaling~\cite{AharonyFisher80}. We surmise that at least two
irrelevant operators must be invoked to account for the above
results---one entering at $\tau^4$, the other at $\tau^6$.

Further remarks on the scaling implications of our work can be found
in section~\ref{sec:comp} while the remainder of the letter will
outline the methods by which the ferromagnetic results were obtained.
A fuller account, including details of the antiferromagnetic
singularity, will appear elsewhere.

\section{Singularity structure and natural boundary}\label{sec:boundary}

It was argued in~\cite{GuttmannEnting96} that on the anisotropic
lattice, the contribution
to the susceptibility of the high-temperature graphs with $2N$
vertical bonds contains more and more poles as $N$ increases, and
that in the limit $N\to\infty$ these poles form a dense set in the
complex plane.  In~\cite{Nickel99} it was shown that in the expansion of the
susceptibility in $j$-particle contributions~\cite{WMTB76}
\begin{equation}
\beta^{-1}\chi=\begin{cases}\sum_{j\text{ odd}}\chi^{(j)} & T>T_c, \\
\sum_{j\text{ even}}\chi^{(j)} & T<T_c, \end{cases}
\label{jparticle}
\end{equation}
the higher-particle components give rise to an ever increasing number of
singularities that appear to form a dense set on the circle $|s|=1$.
In fact, the two phenomena are precisely correlated, with the former
being the highly anisotropic limit, and the latter the isotropic
limit, of the set of singularities for the generic anisotropic model.
These occur at
\begin{align}\label{3.24}
&\cosh(2K)\cosh(2K') - \sinh(2K)\cos{\frac{2\pi m}{j}} \nonumber \\
&\qquad -\sinh(2K')\cos{\frac{2\pi m'}{j}}=0
\end{align}
with $m,m'=1,2,\ldots j,$ and $K,K'=\beta J_x, \beta J_y.$
It will be noted that the
left-hand side of (\ref{3.24}) is the denominator in the Onsager integral
for the free-energy and thus we find the (to us) surprising result
that the singularity of $\chi^{(j)}$, a property of the Ising model in a
magnetic field, is intimately connected with a property in zero
field.
Barring unexpected cancellation (and we have evidence against this)
in the $N\to\infty$ limit, we believe that this set forms a natural boundary.

The existence of a natural boundary has strong implications for our
series analysis. In the $\tau$ plane, the boundary is formed on the
imaginary axis by the logarithmic branch cuts coming from the
singularities of~\cite{Nickel99,Nickel00}.
Summing the contributions of the discontinuities across these cuts
gives $\text{Disc}(\beta^{-1}\chi)\sim\exp(-39.76/{\cal T}^2)$
to leading order, where $\tau = i{\cal T}$ is a point on the cut.
Assuming that the contribution of the physical
singularity at $\tau=0$ is additive, we expect that the coefficient
of $\tau^p$ in the limit $p\to\infty$ in the $\tau$-expansion of the
susceptibility will grow as $ \Gamma(p/2)/a^{p/2} $
where $a=39.76$. This effect only becomes appreciable at order
$p\sim2a$ and is too small to be seen in the terms we have.
However it has the implication that the
$\tau$-expansion of $\chi$ can be at best asymptotic and not
analytic. Whether this non-analyticity comes from the long-distance
or the short-distance part, or a combination of the two, has not
been determined.

We note that the existence of a dense set of singularities also
implies that the susceptibility cannot be a member of the class
of functions called $\cal D$-finite, that is, solutions to linear
differential equations with polynomial coefficients. Nevertheless,
it can be shown from the integral expressions for the $\chi^{(j)}$
that they individually are $\cal D$-finite (also called
holonomic)~\cite{Guttmann00}.
This phenomenon is related to that of perturbative expansions in
quantum field theory, where any individual Feynman diagram is
holonomic whereas the summed series may not be~\cite{KashiwaraKawai77}.

\section{Computation of short-distance amplitudes and
high- and low-temperature series} \label{sec:back}

The essential tool for the computation of both the background
amplitudes and the high- and low-temperature series coefficients is
the set of nonlinear partial difference equations for the two-point
correlation functions, $C(m,n)=\langle\sigma_{00}\sigma_{mn}\rangle$,
given in~\cite{Perk80}. These completely determine all the 
off-diagonal two-point functions once the diagonal ones ($m=n$) are
given. The latter can be computed either by means of an independent
set of difference equations~\cite{JimboMiwa80}
or, as we have done here, directly from the Toeplitz determinant
expressions. The susceptibility, $\beta^{-1}\chi=\sum C(m,n)- 
\langle\sigma_{00}\rangle^2$, is computed by successively adding the
contributions of pairs of square shells $C_N=\sum C(m,n)$ with
$|m|+|n|=2N$ and $|m|+|n|=2N+1$.

The implementation of the difference equations to obtain high- and
low-temperature expansions is straightforward using the multiple
precision integer arithmetic capabilities of Maple$^{\text{TM}}$
 or Mathematica$^{\text{TM}},$
and the time complexity is no worse than ${\rm O}(N^6)$.

The key to computing the short-distance background amplitudes is to
obtain expansions of the partial sums $S_N=\sum_{n=0}^N C_n$
in $\tau$ directly and to identify which terms in the series
contribute to the short-distance part and which to the long-distance
part. A combination of analytic work and numerical fitting leads
us to a conjecture for the short-distance expansion of the shell
sums, namely
\begin{equation}\label{7g}
\sqrt{s} C_N =N^{3/4}\sum_{p=0}^\infty
(\ln|N\tau|)^p(N\tau)^{p^2}A_N^{(p)} 
\end{equation}
where the $A_N^{(p)}$ are Taylor series in $\tau$ with coefficients
that are asymptotic Laurent series in $N^{-1};$ the highest power of
$N$ multiplying $\tau^q$ in $A_N^{(p)}$ is $N^q.$ The partial sums
$S_N$ are
\begin{equation}\label{7n}
\sqrt{s} S_N = \sum^N_{n=0} C_n = \sum_{p=q^2} \sum_{q=0} R_N^{(p,q)}
\tau^p (\ln{|\tau|})^q
\end{equation}
with $R_N^{(p,q)}$ functions of $N$ only. Asymptotically, for large $N$,
$R_N^{(p,q)}$ is a sum of powers $N^{7/4+p'},$ with possible
multiplicative $\ln(N)$ corrections, plus a constant $b^{(p,q)}$
which arises from the small $n$ terms in the sum (\ref{7n}) where the
asymptotic expressions are not valid and sum and integral are not
synonymous. The $p'$ are integers $p'\le p$.

We must assume that (\ref{7n}) remains valid up to $N$ of the order
$1/\tau$ where it can, in principle, be matched term by term to a
large distance expansion that properly describes the roughly
exponential $\exp(-N\tau)$ decay of correlations as $N\rightarrow\infty.$ 
Explicit matching formed the basis of the previous calculations of terms in
the short-distance $\chi$ (cf.\
\cite{KongAuYangPerk86a,Kong87}) but this becomes
extremely cumbersome at higher order. Here we argue that the
exponential decay implies a cutoff on $N$ that is proportional to
$1/\tau$ and that we can identify the temperature behaviour of terms
in $S_N$ in eqn.~(\ref{7n}) by the replacement $N\rightarrow1/\tau.$ All
terms whose variation is as a fractional power of
$\tau$, with possibly logarithmic multipliers, are discarded as
assumed contributions to the long-distance part of $\chi.$ Clearly all
that remains is the constant part of $R_N^{(p,q)}$, namely
$b^{(p,q)}$, and this is extracted by numerical fitting to give
\begin{equation}\label{bkgnd}
\sqrt{s} B = \sum_{p=q^2} \sum_{q=0} b^{(p,q)} \tau^p
(\ln{|\tau|})^q
\end{equation}
for the short-distance part of $\chi$ in eqn.~(\ref{introequation}). 
The coefficients $b^{(p,q)}$ must be determined to very high accuracy
to be useful for the subtraction procedure described in the next
section; the complete list for $p<15$ will be given elsewhere. The
result $p\ge q^2$ we call the fermionic constraint since it can
be traced back to the Toeplitz determinant that led to the
correlations of the form in eqn.~(\ref{7g}).

\section{Scaling amplitudes}\label{sec:scaling}

The contribution of the short-distance terms may now be subtracted
from the high- and low-temperature series, leaving the long-distance part,
from which the scaling amplitudes may be computed using any of a
variety of series analysis techniques. Such analysis is vastly
simplified by the observation that there are no logarithmic or
non-integer power contributions to the scaling-amplitude functions $F_{\pm}$.

To show this, independently of any fitting procedure, we have noted
that any contribution to $F_{\pm}$ which is not a positive integer
power of $\tau$ would manifest itself in the high order series
coefficients of the scaled susceptibility,
$(1-s^{\pm4})^{-1/4}\chi_{\pm}$. The $1+\tau/2$ terms in $F_{\pm}$,
as poles in the scaled susceptibility, also contribute but as their
amplitudes are known to high precision, they can be subtracted. The
residual coefficients are comparable in magnitude to those expected from
the first neglected short-distance term which enters at $\tau^{15}.$
We may place
limits on the size of the amplitudes of any putative non-analytic
terms in the scaling-amplitude functions. For example, for terms of the form
$A_p \tau^p \ln|\tau|$, the bounds,
\begin{eqnarray}\label{8b}
|A_p| & < &  10^{-35} 300^{p}/\Gamma(p-1), \mbox{    }   T>T_c,\\
|A_p| & < &  10^{-37} 600^{p}/\Gamma(p-1), \mbox{     }  T<T_c,
\end{eqnarray}
reasonably exclude all $p$ less than about $15$.

On purely numerical grounds, the absence of logarithmic corrections is
surprising since it implies the
cancellation of the many logarithmic multipliers in the scaling terms
we discarded in the previous section. On the other
hand, the absence of logarithms appears to be a requirement of the
combination of the fermionic constraint $p\ge q^2$ in (\ref{bkgnd}) and
renormalization-group scaling as discussed in the next section.

To compute the amplitudes of the integer
powers of $\tau$ in the scaling-amplitude functions $F_{\pm}$, we have carried
out two independent analyses, one in the $s$-plane, the other in the
$v$-plane, where $v=\tanh K$ is the conventional high-temperature
variable. The natural boundary singularities at $|s|=1$ are mapped
to two circles, $|v\pm1|=\sqrt{2}$. As they are farther from the
origin than the ferromagnetic and antiferromagnetic singularities at
$v=\pm(\sqrt{2}-1)$ their
amplitudes are exponentially damped and may be neglected in the
analysis. The $s$-plane analysis must take account of these
singularities explicitly.  The two analyses are in complete
agreement.

We find numerically that the scaling-amplitude functions multiplied by
$\sqrt{s}$ appear to be even functions of $\tau$, the amplitudes of the odd
terms being comparable in magnitude to the uncertainties in the even
amplitudes. 
The rational amplitudes of $\tau^2$ and $\tau^4$ below we conjecture to
be exact, and these values were fixed in the final fitting.
The results, with
uncertainty only in the final digits, are
\begin{align}\label{results}
\sqrt{s}F_+ & =1 +\tau^2/2 -\tau^4/12 -0.1235292285752086663 \tau^6\nonumber\\
& +0.136610949809095 \tau^8 -0.13043897213 \tau^{10}\nonumber\\
& +0.1215129 \tau^{12} -0.113 \tau^{14} + {\rm O}(\tau^{15}),\nonumber\\
\sqrt{s}F_- & = 1 +\tau^2/2 -\tau^4/12\nonumber\\
&  -6.321306840495936623067\tau^6\nonumber\\
& +6.25199747046024329 \tau^8  - 5.6896599756180 \tau^{10} \nonumber\\
& +5.142218271 \tau^{12} -4.67472 \tau^{14} + {\rm O}(\tau^{15}).
\end{align}

\section{Comparison with scaling predictions}\label{sec:comp}

Prior to the analysis of~\cite{Nickel00}, all known amplitudes
were in agreement with the hypothesis that corrections to scaling
were due to scaling-field nonlinearity, and not to the presence of
irrelevant variables. Here for the first time, we quantify the error
in this ``simple'' scaling theory. Ignoring irrelevant operators, the
expressions for the free energy, magnetization and susceptibility in
zero magnetic field are~\cite{AharonyFisher80}
\begin{align}\label{s3}
f(\tau) & = - A (a_0(\tau))^2\ln{|a_0(\tau)|} + A_0(\tau),
\nonumber\\
M(\tau<0) & = B b_1(\tau)|a_0(\tau)|^{1/8},  \nonumber \\
\beta^{-1}\chi_{\pm}(\tau) & = C_{\pm}(b_1(\tau))^2|a_0(\tau)|^{-7/4}
\nonumber\\
 & - E a_2(\tau)a_0(\tau)\ln{|a_0(\tau)|} + D(\tau),
\end{align}
where $A$, $B$, $C_{\pm}$, and $E$ are constants and $A_0(\tau)$ and
$D(\tau)$ are analytic functions of $\tau$. The functions $a_0(\tau)$ 
and $b_1(\tau)$ are the leading terms in the expansion of the scaling 
fields, and can be determined from the free energy and magnetization. 
The result for $\chi_{\pm}$ is of the form (\ref{introequation}) but 
with $F$ replacing $F_{\pm}$ where 
\begin{equation}\label{9c}
\sqrt{s}F= 1 +\frac{\tau^2}{2} -\frac{31\tau^4}{384}
 +\frac{125\tau^6}{3072} + {\rm O}(\tau^8).
\end{equation}
Note that this expression should hold in both temperature regimes.

The difference between~(\ref{9c}) and~(\ref{results}) we believe to
be due to the effects of one or more irrelevant operators, confluent
with the ``simple'' scaling terms.  As there
is no free parameter to vary in the model, we can't identify these
operators from the information we have. However, it is likely that
there are at least two mechanisms at work, one entering at ${\rm
O}(\tau^4)$ which preserves the equality of $F_+$ and $F_-$, and a
second entering at ${\rm O}(\tau^6)$ which breaks this symmetry. In
order to probe these effects further, we hope to study
the model with anisotropy, and on other periodic lattices.

The corrections to scaling we have found are confluent with expected
analytic terms and in the renormalization group picture of scaling this
leads to the possibility of logarithmic terms as well (cf.~\cite{Wegner76}).
Logarithmic corrections are not demanded---the issue is whether the
scaling fields are coupled
and this depends on microscopic details.  Barma and
Fisher~\cite{BarmaFisher84} have
investigated a model renormalization group flow in detail and conclude
that in the case of a confluence, here labelled by integer index $m$, one
must expect either no coupling between fields or corrections of the form
$(\tau^m\log|\tau|)^k$ to all order $k$.  Since the latter violates
the fermionic constraint $m k\ge k^2$  we
conclude there cannot be {\bf any} logarithmic terms in the
scaling-amplitude function $F_{\pm}$ as we have verified to O$(\tau^{15})$.
\newline

\begin{acknowledgments}

We are pleased to acknowledge M. Bousquet-M\'elou, M. E. Fisher,
M. L. Glasser, B. M. McCoy, A. Pelissetto and A. D. Sokal for helpful
comments and criticisms.
AJG and WPO would like to thank the Australian Research Council
for financial support, and JHHP thanks the NSF for support in part
by Grant PHY 97-22159.
\end{acknowledgments}


\end{document}